
\input epsf

\input harvmac

\def\figin{\epsfcheck\figin}\def\figins{\epsfcheck\figins}
\def\epsfcheck{\ifx\epsfbox\UnDeFiNeD
\message{(NO epsf.tex, FIGURES WILL BE IGNORED)}
\gdef\figin##1{\vskip2in}\gdef\figins##1{\hskip.5in}
\else\message{(FIGURES WILL BE INCLUDED)}%
\gdef\figin##1{##1}\gdef\figins##1{##1}\fi}
\def\DefWarn#1{}
\def\figinsert{\goodbreak\midinsert}
\def\ifig#1#2#3{\DefWarn#1\xdef#1{fig.~\the\figno}
\writedef{#1\leftbracket fig.\noexpand~\the\figno}%
\figinsert\figin{\centerline{#3}}\medskip\centerline{\vbox{\baselineskip12pt
\centerline{\footnotefont{\bf Fig.~\the\figno:} #2}}}
\bigskip\endinsert\global\advance\figno by1}

\def\Title#1#2{\rightline{#1}\ifx\answ\bigans\nopagenumbers\pageno0
\vskip0.5in
\else\pageno1\vskip.5in\fi \centerline{\titlefont #2}\vskip .3in}

\font\caps=cmcsc10

\noblackbox
\parskip=1.5mm


\def\npb#1#2#3{{\it Nucl. Phys.} {\bf B#1} (#2) #3 }
\def\plb#1#2#3{{\it Phys. Lett.} {\bf B#1} (#2) #3 }
\def\prd#1#2#3{{\it Phys. Rev. } {\bf D#1} (#2) #3 }
\def\prl#1#2#3{{\it Phys. Rev. Lett.} {\bf #1} (#2) #3 }
\def\mpla#1#2#3{{\it Mod. Phys. Lett.} {\bf A#1} (#2) #3 }
\def\ijmpa#1#2#3{{\it Int. J. Mod. Phys.} {\bf A#1} (#2) #3 }
\def\jmp#1#2#3{{\it J. Math. Phys.} {\bf #1} (#2) #3 }
\def\cmp#1#2#3{{\it Commun. Math. Phys.} {\bf #1} (#2) #3 }
\def\pr#1#2#3{{\it Phys. Rev. } {\bf #1} (#2) #3 }
\def\umn#1#2#3{{\it Usp. Matem. Nauk} {\bf #1} (#2) #3 }
\def\bb#1{{\tt hep-th/#1}}


\def\dj{\hbox{d\kern-0.347em \vrule width 0.3em height 1.252ex depth
-1.21ex \kern 0.051em}}

\def\gst{g_{\rm st}}
\def\half{{1\over 2}\,}
\def\d{{\rm d}}
\def\ee{{\rm e}\,}
\def\Re{{\rm Re\,}}
\def\Im{{\rm Im\,}}
\def\Tr{{\rm Tr\,}}

\def\ket{\rangle}
\def\bra{\langle}

\def\sign{{\rm{sign}}}


           \def\CO{{\cal O}}


\def\ap{\alpha_{+}}
\def\am{\alpha_{-}}
\def\ta{\tilde\alpha}
\def\tap{\tilde\alpha_{+}}
\def\tam{\tilde\alpha_{-}}
\def\tapm{\tilde\alpha_{\pm}}
\def\tamp{\tilde\alpha_{\mp}}
\def\ttap{\hat\alpha_{+}}
\def\ttam{\hat\alpha_{-}}
\def\ttapm{\hat\alpha_{\pm}}

\def\ttcp{\hat c^{+}}
\def\ttcm{\hat c^{-}}
\def\ttcpm{\hat c^{\pm}}

\def\bp{\beta_{+}}
\def\bm{\beta_{-}}
\def\apm{\alpha_{\pm}}

\def\xmes{{\d x\over 2\pi}\,}
\def\pmes{{\d p\over 2\pi}\,}
\def\taumes{{\d \tau \over 2\pi}\,}

\def\PB{\{\apm(x), \apm(y)\} = \mp 2\pi \delta'(x-y) }

\def\intinf{\int_{-\infty}^{\infty}\,}
\def\inthinf{\int_0^{\infty}\,}
\def\mf{\mu_{\rm F}}
\def\delx{\partial_x}
\def\delt{\partial_t}
\def\deltau{\partial_\tau}
\def\pp{\pi\phi_0}
\def\ppx{\pi\phi_0(x)}
\def\pptau{\pi\phi_0(\tau)}
\def\tpp{T^{(+)}_{ik,+}}
\def\tpm{T^{(+)}_{ik,-}}
\def\tp{T^{(+)}_{ik}}

\def\da{\d\alpha\,\,}
\def\mx{{M\over x^2}\,}
\def\intamp{\int_{\am}^{\ap}\,}
\def\intap{\int_{0}^{\ap}\,}
\def\intam{\int_{0}^{\am}\,}
\def\intamp{\int_{\am}^{\ap}\,}
\def\ak{\alpha_k}
\def\akp{\alpha_{k'}}
\def\LpM{\Lambda_+ + {M\over x^2}\,}

\def\LpmM{\Lambda_{\pm} + {M\over x^2}\,}
\def\Lppp{{\Lambda_+\over \pp}\,}

\def\Lpmpp{{\Lambda_{\pm}\over \pp}\,}
\def\ektau{\ee^{ik\tau}\,}
\def\nn{n_{\rm N}}
\def\sn{\Sigma_{\rm N}}
\def\snone{\Sigma_{{\rm N}-1}}
\def\pikf{\bigl({\pi\over 4}\,k\bigr)}

\def\kn{k_{{\rm 2N}}}


\lref\rGeDi{I.~M. Gelfand and L.~A. Dikii, \umn {30}{1975}{67.}}
\lref\rDaMa{R. Dashen, S-K. Ma and H.~J. Bernstein,
\pr {187}{1969}{345.}}
\lref\rGR{I.~S. Gradshteyn and I.~M. Ryzhik,
{\it Table of integrals, series, and products},
(Academic Press, 1980).}
\lref\rREVi{
I. R. Klebanov, ``{\it String theory in two dimensions},''
in ``String Theory and Quantum Gravity'',  Proceedings of the Trieste
Spring School 1991, eds. J. Harvey et al., (World Scientific,
Singapore, 1992)\semi
P. Ginsparg and G. Moore, ``{\it Lectures on 2D string theory},''
Lectures given at 1992 TASI summer school, YCTP-P23-92 and
LA-UR-92-3479, \bb {9304011}.}
\lref\rREVii{
A. Jevicki, ``{\it Fields and symmetries od 2d strings},''
Lectures at the 7th Nishinomiya--Yukawa Memorial Symposium 1992,
BROWN-HET-894, \bb{9302106}\semi
D. Kutasov, ``{\it Some properties of (non) critcal strings},''
in ``String Theory and Quantum Gravity'',  Proceedings of the Trieste
Spring School 1991, eds. J. Harvey et al., (World Scientific,
Singapore, 1992).}
\lref\rWi{E. Witten, \prd {44} {1991} {314.}}
\lref\rBH{G. Mandal, A. Sengupta and S. Wadia, \mpla {6}{1991}{1685\semi}
I. Bars and D. Nemeschansky, \npb {348}{1991}{89\semi}
S. Elizur, A. Forge and E. Rabinovici, \npb {359}{1991}{581\semi}
M. Ro\v cek, K. Schoutens and A. Sevrin, \plb {265}{1991}{303.}}
\lref\rDVV{R. Dijkgraaf, E. Verlinde and H. Verlinde,
\npb {371} {1992} {269.}}
\lref\rBHmm{
S.~R. Das, \mpla {8} {1993} {69\semi}
A. Dhar, G. Mandal and S. Wadia, \mpla {7}{1992}{3703\semi}
J.~G. Russo, \plb {300}{1993}{336\semi}
Z. Yang, ``{\it A possible black hole background in $c=1$ matrix
model},'' UR-1251, \bb{9202078}.}
\lref\rJeYo{A. Jevicki and T. Yoneya, ``{\it A deformed matrix model and
the black hole background in two-dimensional string theory},''
NSF-ITP-93-67, BROWN-HEP-904, UT-KOMABA/93-10, \bb{9305109}.}
\lref\rDa{U. Danielsson, ``{\it A matrix-model black hole},''
CERN-TH.6916/93, \bb{9306063}.}
\lref\rAvJe{J. Avan and A. Jevicki, \cmp {150} {1992} {149.}}
\lref\rAvJeold{J. Avan and A. Jevicki, \plb {266}{1991}{35;}
\plb {272}{1991}{17.} }
\lref\rLa{F. Calogero, \jmp {10} {1969} {2191\semi}
F. Zirilli, \jmp {15} {1974} {1202\semi}
L. Lathouwers, \jmp {16} {1975} {1393.} }
\lref\rGrMi{D.~J. Gross and N. Miljkovi\'c, \plb {238} {1990} {217.}}
\lref\rCFT{ A. Jevicki and B. Sakita, \npb {165} {1980} {511\semi}
S.~R. Das and A. Jevicki, \mpla {5} {1990} {1639.}}
\lref\rJeSa{A. Jevicki and B. Sakita, \npb {165} {1980} {511.}}
\lref\rDaJe{S.~R. Das and A. Jevicki, \mpla {5} {1990} {1639.}}
\lref\rDJR{ K. Demeterfi, A. Jevicki and J.~P. Rodrigues, \npb {362}{1991}
{173;}  \npb {365} {1991} {499;} \mpla {6} {1991} {3199.}}
\lref\rJRT{A. Jevicki, J.~P. Rodrigues and A. van Tonder,
``{\it Scattering states and symmetries in the matrix model
and two dimensional string theory},'' BROWN-HET-874 and
CNLS-92-10 ({\it Nucl. Phys.} {\bf B} in press),
\bb{9209057}.}
\lref\rMoSe{G. Moore and N. Seiberg, \ijmpa{7}{1992}{2601.}}
\lref\rMoPl{G. Moore and R. Plesser, \prd {46} {1992} {1730.}}
\lref\rMoore{G. Moore, \npb {368} {1992} {557.}}
\lref\rJoao{J.~P. Rodrigues, unpublished.}
\lref\rCFTampl{
D.~J. Gross and I.~R. Klebanov, \npb {356} {1991} {3\semi}
P. Di Francesco and D. Kutasov, \plb {261} {1991} {385;}
\npb {375} {1991} {119.}}
\lref\rGKN{D.~J. Gross, I.~R. Klebanov and M.~J. Newman,
\npb {350} {1991} {621.}}
\lref\rPo{ J. Polchinski, \npb {362} {1991} {125.}}
\lref\rGrDa{U.~H. Danielsson and D.~J. Gross,\npb{366}{1992}{3\semi}
U.~H. Danielsson, \npb {380}{1992}{83.}}
\lref\rBeKl{M. Bershadsky and I.~R. Klebanov, \prl {65}{1990}{3088;}
\npb {360}{1991}{559\semi}
N. Sakai and Y. Tanii, \ijmpa {6}{1991}{2743.} }
\lref\rGrKlcircle{D.~J. Gross and I.~R. Klebanov,
\npb {344}{1990}{475.}}



\Title{\vbox{\baselineskip 12pt\hbox{PUPT-1407}\hbox{CNLS-93-06}
\hbox{{\tt hep-th/9306141}} }}
{\vbox {\centerline{States and Quantum Effects in the}
\medskip
\centerline{Collective Field Theory of a}
\medskip
\centerline{Deformed Matrix Model}  }}

\centerline{$\quad$ {\caps Kre\v simir Demeterfi}\footnote{$^\dagger$}
{On leave of absence from the Ru\dj er Bo\v skovi\'c Institute,
Zagreb, Croatia}}
\smallskip
\centerline{{\sl Joseph Henry Laboratories}}
\centerline{{\sl Princeton University}}
\centerline{{\sl Princeton, NJ 08544, U.S.A.}}
\centerline{{\tt kresimir@puhep1.princeton.edu}}
\medskip
\centerline{and}
\medskip
\centerline{{\caps Jo\~ao P. Rodrigues}}
\smallskip
\centerline{{\sl Physics Department and Centre for Nonlinear Studies}}
\centerline{{\sl University of the Witwatersrand}}
\centerline{{\sl Wits 2050, South Africa}}
\centerline{{\tt 005rod@witsvma.wits.ac.za}}
\vskip 0.4in

We derive an equation which gives the tree-level
scattering amplitudes for tachyons in the black
hole background using the exact states of the collective
field hamiltonian corresponding to a deformed matrix model
recently proposed by Jevicki and Yoneya.
Using directly the symmetry algebra we obtain
explicit expression for a class of amplitudes in
the tree approximation.
We also study the quantum effects in the corresponding
collective field theory.
In particular, we compute the ground state energy
and the free energy at finite temperature
up to two loops, and the first quantum correction
to the two-point function.


\Date{6/93}


\newsec{Introduction}

Matrix models have been a very powerful tool for
studying two-dimensional quantum gravity and low-dimensional string
theories.
Although there are still
some puzzles in establishing the exact
relationship between the matrix model and the standard
continuum formulation of string theory, it is
generally believed that the one-dimensional hermitian
matrix model in the double scaling limit describes
the linear dilaton solution  with tachyon condensation of
the two-dimensional critical string theory in a flat
space-time.  The main reason for this belief is
an exact agreement of all known results  for the
correlation functions
and the existence of the $W_\infty$ symmetry in both
approaches (for reviews see e.g.\refs{\rREVi,\rREVii}).

If, however, the matrix model is to be taken as a
(nonperturbative) definition of  string theory
it must in some way encompass {\it all} solutions of
string theory, in particular the
black hole background\refs{\rWi\rBH{--}\rDVV}.
The importance of a full quantum mechanical
(and possibly nonperturbative) understanding of
black hole physics can hardly be overemphasized.
Since the matrix model, at the present time, provides
the most elegant framework for explicit computations it
is very important to find a matrix model formulation of the
black hole background.  Despite numerous attempts\refs{\rBHmm}
the problem remains unsolved.
An important step in this direction
has recently  been taken by Jevicki and Yoneya\refs{\rJeYo}.

Their proposal is based on several lessons we have
learned from the standard $c=1$ model. First, the massless
scalar collective field\refs{\rCFT} is related
to the string theory tachyon  through a
non-local redefinition\refs{\rMoSe} which transforms
the Klein-Gordon equation of the collective field theory into the
Virasoro condition in the string theory. It is remarkable that
the wave function renormalization induced by this
transformation precisely accounts for the
``external leg factors'' in the tachyon scattering
amplitudes which have the following factorized form\refs{\rCFTampl}:
$$A\,({\rm tachyons}\,)= ({\rm external\,\,\,leg\,\,\,factors}\,)
\times A\,({\rm collective\,\, field\,\,\, theory})\,\,.$$
Although in Minkowski space external leg factors are pure
phases and therefore have no physical effect they still
contain physical information about the background.
Namely, after analytic continuation to Euclidean space
they have poles at special discrete values of momenta
for which the incoming tachyon wave is in resonance with
the ``wall'' condensate.
The on-shell collective field amplitudes, on the other hand,
are simple polynomials in certain  combinations of the
(absolute values) of momenta and have no poles\refs{\rDJR}.

Another crucial point in the proposal\refs{\rJeYo} is
related to the identification of the string coupling
constant $\gst$. In the standard $c=1$ case it is
given by $\gst \sim 1/\mu$, where $\mu$ is the (negative)
Fermi energy which can be thought of as a constant deformation
of the inverted harmonic potential in the corresponding
matrix model.

In the black hole background,
the non-local field redefinition of the collective field
which relates it to the tachyon field was discussed in\refs{\rBHmm}.
It is given a new interpretation in\refs{\rJeYo}
where it is also argued the that the existence of such
transformation is necessary, but not sufficient
for the matrix model to describe the black hole
background.  In addition one must have
the correct relationship
between the string coupling constant and the black hole
mass, $\gst \sim 1/\sqrt{M}$, which can be achieved if one deformes
the inverted harmonic oscillator potential
by a singular term $M/x^2$ and sets the Fermi energy to
zero\refs{\rJeYo}.
This is precisely the deformation of the inverted harmonic
potential found by Avan and Jevicki\refs{\rAvJe} in a search
for potentials with algebraic structures similar
to the one found in the standard $c=1$ case\refs{\rAvJeold}.

It is also shown by Jevicki and Yoneya that the
factorization assumption is consistent only
if the amplitudes for odd number of tachyons vanish.
This imposes strong constraints on the form of the
collective field theory, and they
demonstrated that the deformed potential
indeed gives scattering amplitudes
which satisfy this constraint (at least
in the tree approximation).
Therefore, the model proposed in\refs{\rJeYo}
is the first matrix model consistent with minimal
requirements to make connection with the black hole
background and deserves a careful analysis.
In this article we study further the properties of the
collective field formulation of the deformed matrix model
concentrating on states, symmetries, quantum and
finite temperature effects.

This paper is organized as follows:
In sect.~2 we review the collective field formulation
of a general matrix model including normal ordering and apply it to
the deformed matrix model.
In sect.~3 we summarize some results
on the exact eigenstates and symmetries
of the collective hamiltonian\refs{\rAvJeold} which are needed
for our discussion.
Following ref.\refs{\rJRT} we then derive an equation which describes
the tree-level scattering amplitudes using these exact states.
In sect.~4 we give another derivation of scattering
amplitudes using directly the symmetry algebra.
In sect.~5 we present some examples of loop calculations.
In particular, we compute the ground state energy
and the finite temperature free energy up to two loops
and the two-point function up to one loop.
We find that the free energy is not invariant under the
duality transformation $2\pi T \to 1/(2\pi T)$.
Sect.~6 is reserved for discussion and conclusions.
In appendix A we derive some of the results of
sect.~5 directly in the fermionic picture of the
matrix model in order to confirm some of our results and to provide
further evidence of the completeness (at least perturbatively)
of the normal ordered collective hamiltonian.

\newsec{Collective field theory of a deformed matrix model}

Let us consider a general one-dimensional hermitian matrix
model defined by
\eqn\Emmh
{H = \Tr \Bigl(\half \dot M^2(t) + V(M) \Bigr)\,\,.}
The singlet sector of this hamiltonian describes the
dynamics of $N$ decoupled nonrelativistic fermions in
an external potential with the single particle hamiltonian
given by
\eqn\Esph
{h(p,x) = \half p^2 + v(x) - \mf \,\,,}
where $\mf$ is the Fermi energy.
The collective field theory\refs{\rCFT}
is most easily derived in the Fermi liquid picture
first discussed by Polchinski\refs{\rPo} with the
result
\eqn\Ecompact
{H =\intinf \xmes\int_{\am(x)}^{\ap(x)}\,\da \,\,h(\alpha,x)\,\,.}
As noted in\refs{\rJeYo} one could equally take the collective
hamiltonian to be
\eqn\Echpspace{
H = \intinf \pmes \int_{\bm(p)}^{\bp(p)} \, \d\beta \,\,h(p,\beta)\,\,,}
which would lead to a differently looking theory but
which should give the same physical results.
In the following we will discuss only the first representation.

Integrating \Ecompact\ one finds
\eqn\Egc
{H = \intinf \xmes \Bigl[{1\over 6}\,\bigl( \ap^3 - \am^3 \bigr) -
\bigl( \mf - v(x) \bigr) \bigl( \ap - \am \bigr) \Bigr]\,\,.}
The fields $\apm(x,t) \equiv \delx\Pi \pm \pi\phi$
represent linear
combinations of the canonical field and momentum, and
can be identified with the two branches of the
profile function of the Fermi surface in the
classical phase space.
They obey  Poisson brackets $\PB$.
The classical equation of motion
$\delt \apm = \{\apm, H\}$ has a static solution
which is easily found to be
\eqn\Eclsol
{\pi\phi_0(x) = \sqrt{2[\mf - v(x)]}\,\,
\theta[\mf - v(x)] \,\,.}

Changing variables from $x$ to the ``time-of-flight'' variable $\tau$
\eqn\Etaux
{\d\tau= {\d x \over \ppx}\,\,,}
and shifting the fields $\apm(x,t)$ by the classical background
$\pi\phi_0$:
\eqn\Eshift
{\apm(x,t) = \pm\pp + {1\over \pp}\,\tapm(\tau,t)\,\,,}
one obtains the following hamiltonian
\eqn\Echtau
{H = \half\inthinf\taumes \Bigl(\tap^2 + \tam^2 \Bigr) +
{1\over 6}\,\inthinf \taumes {1\over (\pp)^2}\,\,
\Bigl( \tap^3 -  \tam^3 \Bigr) + E^{(0)}\,\,,}
where the constant $E^{(0)}$
is the classical energy of the background field
\eqn\Eezerodef
{E^{(0)} = -{1\over 3\pi}\,\int \d x\,\,[2(\mf - v(x))]^{3/2}\,\,.}
In the quadratic part of the hamiltonian  one
easily recognizes a massless two-dimensional scalar theory
since $\tapm(\tau) = -\Pi_\zeta \pm \pi\deltau\zeta$.
This is true for
{\it any} matrix model provided there is a static classical
solution. The coupling constant of the cubic interaction
is space dependent and encodes the specific form of the matrix
model potential through $\pp$.

At the linearized level, $\tapm$ are simply the
right and left moving massless modes which satisfy
$(\delt \pm \deltau)\,\tapm(\tau,t) = 0$, and can be
expanded as
\eqn\Eaexp{
\tapm(\tau,t)=\pm\intinf \d k\,\,\ak^{\pm}\,\ee^{-ik(t\mp\tau)}\,,\quad
[\ak^{\pm},\akp^{\pm}]=k\delta(k+k')\,,\quad
[\ak^{\pm},\akp^{\mp}]=0\,\,.}
Due to the Dirichlet boundary conditions for the shifted
field, the ``plus'' and ``minus'' sectors of the theory
\Echtau\ are not independent.
It was argued in ref.\refs{\rJRT}
that the definition of the $\tapm$ fields can be extended to
the whole real line by requiring $\tapm (-\tau)= - \tamp (\tau)$,
which is equivalent to
\eqn\Ebc{ \ak^{\mp} = -\ak^{\pm} \,\,.}
In physical terms
this means that creation of a right-mover must be associated
with the creation of a left-mover but with opposite amplitude
(and similarly for the annihilation processes).
In the standard perturbative expansion of the collective
field theory, this condition is explicitly
built into the field expansions,
and if one chooses $\tap$ as the independent
field (as we will choose to do in the following), one finds that
all contributions from the ``minus'' sector
of the theory always extend the range of $\tau$ integration of the
``plus'' sector to the whole real line. Thus the
hamiltonian \Echtau\ can be written as
\eqn\Ehcf
{H=\half \intinf\taumes \alpha^2(\tau)+
{1\over 6}\,\intinf\taumes {1\over (\pp)^2}\,\alpha^3(\tau)\,\,.}
To simplify notation we have dropped the subscript
$+$ and the tilde.
We will see later that condition \Ebc, when imposed on
exact states expressed in terms of asymptotic variable
expansions, determines all tree-level scattering amplitudes.

The full quantum hamiltonian is obtained by normal
ordering\refs{\rDJR}, by taking into account the fact that
the hamiltonian \Ehcf\ was obtained following
a reparametrization from the $x$ to the $\tau$ coordinate. Denoting
normal ordering in $x$-space by
$$N_x(\alpha^2) = \alpha^2 - \bra\alpha\alpha\ket_x \,\,,$$
and in $\tau$-space by
$$N_\tau(\alpha^2) = \alpha^2 - \bra\alpha\alpha\ket_\tau \,\,,$$
we have
$$ N_\tau(\alpha^2) = N_x(\alpha^2) + \Delta(\tau)\,\,.$$
$\Delta(\tau)$ is simply the (finite) difference between the two
propagators and it can be expressed solely in terms of classical
solution $\phi_0$ and its derivatives as follows:
\eqn\EDtau
{\Delta(\tau)={1\over12}\,\Bigl[2\,\Bigl({\phi_{0}''\over \phi_0}\Bigr)
-3\,\Bigl({\phi_{0}'\over\phi_0}\bigr)^2\,\Bigr] \,\,.}
For an arbitrary polynomial in $\alpha$ one has the
following relation between normal ordering in the two spaces
\eqn\Eno
{N_x[P(\alpha)] = N_\tau \left( \ee^{\,\half\Delta(\tau)\,
{\partial^2\over\partial\alpha^2} }\,\, P(\alpha) \,\right)\,\,.}
The normal ordered (in $x$-space) hamiltonian \Ehcf\ reads
\eqn\Ehno
{H =\half\intinf\taumes :\alpha^2:
+{1\over6}\,\intinf\taumes {1\over (\pp)^2}\,:\alpha^3:
+\half\intinf\taumes {\Delta(\tau)\over (\pp)^2}\,:\alpha:
+E^{(1)} \,\,,}
where double dots denote the normal ordering in $\tau$-space.
The normal ordering introduces a
finite linear tadpole, and the last term in \Ehno\ which is
independent of $\alpha$ is the first quantum correction to the
ground state energy
\eqn\Eeonedef
{E^{(1)}=\half\intinf\taumes \Delta(\tau) \,\,.}
For practical purposes it is convenient to express
the hamiltonian in momentum space. One finds
\eqn\Ehms
{H=\half\intinf \d k\,:\alpha_k \alpha_{-k}: + {1\over6}\,\intinf
\d^3 k\,\, f(k_1+k_2+k_3)
:\alpha_{k_1} \alpha_{k_2} \alpha_{k_3}:
+\half\intinf \d k\,\, g(k)\,\alpha_k }
where
\eqn\Efgdef {\eqalign{
&f(k)=\intinf\taumes f(\tau)\,\ee^{ik\tau}\,,\qquad
f(\tau) = {1\over(\pp)^2}\,\,,\cr
\noalign{\vskip 0.2truecm}
&g(k)=\intinf\taumes g(\tau)\,\ee^{ik\tau}\,,\qquad
g(\tau) = {\Delta(\tau)\over(\pp)^2}\,\,.\cr} }
$\ak$ is creation (annihilation) operator for $k<0$ ($k>0$).

The above description of the collective field theory is perfectly
general and it applies to any matrix model potential.
In what follows we consider the model with zero Fermi energy
($\mf=0$) and with a deformed inverted harmonic potential
\eqn\Edmm
{v(x) = - \half x^2 + {M\over 2x^2}\,\,,}
suggested in\refs{\rJeYo} as a candidate for describing the black
hole background. In this case the classical solution \Eclsol\ reads
\eqn\Ecsx
{\ppx = \sqrt{x^2 - {M\over x^2}}\,\,.}
Integrating eq. \Etaux\ and then inverting $\tau(x)$ one finds
\eqn\Exoftau
{x(\tau) = M^{1/4}\,\sqrt{\cosh 2\tau}\,\,,}
which then gives
\eqn\Ecstau
{\pptau = M^{1/4}\,{\sinh 2\tau \over \sqrt{\cosh 2\tau}}\,\,.}
{}From this one simply finds
\eqn\Ealltau{\eqalign{
&f(\tau) ={1\over\sqrt{M}}\,{\cosh 2\tau\over \sinh^2 2\tau}\,\,,\cr
\noalign{\vskip 0.2truecm}
&\Delta(\tau)={-1\over 12}\,\Bigl(1 + {3\over\cosh^2 2\tau} +
{12 \over\sinh^2 2\tau}\Bigr)\,\,,\cr
\noalign{\vskip 0.2truecm}
&g(\tau)={1\over\sqrt{M}}\,\Bigl({1\over 4\cosh 2\tau} +
{2\cosh 2\tau\over 3\sinh^2 2\tau} -
{\cosh^3 2\tau\over \sinh^4 2\tau} \Bigr)\,\,.\cr} }
After a Fourier transform one gets
\eqn\Efgk{\eqalign{
&f(k)={-k\over 8\sqrt{M}}\,\tanh \pikf \,\,,\cr
\noalign{\vskip 0.2truecm}
&g(k)={1\over 16\sqrt{M}}\,\Bigl[
{1\over \cosh\pikf} + {1\over 3}\,\bigl(k-{k^3\over 4}\bigr)
\tanh\pikf\,\Bigr]\,\,.\cr} }
In performing Fourier transforms,
any poles on the contour of integration
have been handled by means of a principal part prescription,
as it was physically justified in ref.\refs{\rDJR}.

The cubic vertex $f(k)$ was already derived in\refs{\rJeYo}.
It has an interesting property that $f(0)=0$ which is
crucial for the tree-level three- and five-point
amplitudes to vanish on-shell.
We note that both the cubic vertex $f(k)$ and the
linear tadpole $g(k)$ have poles at discrete
imaginary values of momenta $ik = 2(2n+1)$, i.e. when the
momentum is twice an odd integer.
For comparisson, in the standard $c=1$ case
with potential $v(x)=-x^2/2$ vertex functions
are proportional to $\coth(\pi k/2)$, and therefore
have poles when the (imaginary) momentum equals an
even integer, i.e. $ik=2n$.

The cubic vertex $f(k)$ and the linear tadpole $g(k)$ provide
a basis for systematic perturbative computations.
Before presenting some examples of
perturbative calculations using the above hamiltonian
we consider in the next section states and symmetries of
the classical theory and derive the tree-level
scattering equation.

\newsec{Exact states and symmetries}

The symmetry structure of the collective hamiltonian
with the  potential \Edmm\ was described by Avan and
Jevicki\refs{\rAvJe} in terms of three-index operators:
\eqn\Eop
{O_{j,m}^{a} \equiv \int \xmes \intamp \da
\Bigl( \alpha^2 - x^2 +\mx \Bigr)^a
\Bigl( (\alpha + x)^2 +\mx \Bigr)^{{j+m\over 2}}
\Bigl( (\alpha - x)^2 +\mx \Bigr)^{{j-m\over 2}}}
which satisfy the following algebra:
\eqn\Ealg
{\bigl[ O_{j_1,m_1}^{a_1},  O_{j_2,m_2}^{a_2} \bigr] =
-4i(j_1 m_2 - m_1 j_2)\,O_{j_1+j_2-2, m_1+m_2}^{a_1+a_2+1}
-4i(a_1 m_2 - m_1 a_2)\,O_{j_1+j_2, m_1+m_2}^{a_1+a_2-1} \,\,.}
This algebra can be reduced  to a two-index algebra by
means of the identity
\eqn\Ethreetwo
{O_{j,m}^{a+2} = O_{j+2,m}^{a} - 4M O_{j,m}^{a}\,\,.}
The operators \Eop\ satisfy the following commutation relation
with the collective hamiltonian $H$:
\eqn\EcomHO
{ [ H, O_{j,m}^{a} ] = -i\,2m\,O_{j,m}^{a} \,\,.}

In the above, $a$ and $j$ are integers and $m=-j,-j+2,...,j-2,j$.
This choice has been made to ensure that
the exponents in \Eop\ are always integers which then guarantees
that the operators $O_{j,m}^{a}$ are polynomial eigenstates
(i.e., they have finite expansions)
in the sense of refs.\refs{\rAvJe,\rAvJeold}.
It then follows from \EcomHO\
that the energies of the discrete states of the theory
are (imaginary) {\it even} numbers, in agreement
with other independent considerations\refs{\rDVV,\rJeYo}.
A consequence of this choice is that in the limit
${M \to 0}$ one recovers only ``half'' the $W_{\infty}$ algebra of
the $c=1$ model. This is very precisely related to the choice of
physically acceptable solutions of the  Schr\"odinger equation
associated with the potential \Edmm\ as it will be
discussed in appendix A.

Special cases of the operators \Eop\ which will be useful
in our future discussion are:
\eqn\EOTp {\eqalign{
&O_{j,j}^{a=0} = \int \xmes \intamp \da
\Bigl[ (\alpha +x)^2 + \mx\Bigr]^j \equiv T_{2j}^{(+)}\,\,, \cr
\noalign{\vskip 0.2truecm}
&O_{j,-j}^{a=0} = \int \xmes \intamp \da
\Bigl[ (\alpha - x)^2 + \mx\Bigr]^j \equiv T_{2j}^{(-)}\,\,, \cr
\noalign{\vskip 0.2truecm}
&O_{0,0}^{a=1} = \int \xmes \intamp \da
\Bigl[\alpha^2 - x^2 + \mx\Bigr] \quad\equiv 2H \,\,. \cr } }

In order to discuss tachyon amplitudes one needs to
make an analytic continuation of the operators \Eop\ and
the algebra \Ealg. This procedure was justified
for the  standard $c=1$ model in ref.\refs{\rJRT}.
Here we will argue that the same is true in the case
of a deformed matrix model.
By letting $j \to \pm ik/2$ one obtains from \EOTp\
operators $T^{(-)}_{-ik}$ ($\tp$) which when acting
on vacuum create an in (out) state, as we now demonstrate.
Consider for example the operator
\eqn\Etpdef
{\tp = \int \xmes \intamp \da
\Bigl[ (\alpha +x)^2 + \mx \Bigr]^{ik/2}\,\,.}
Shifting $\alpha(x)$  as in \Eshift\ and expanding it up to
quadratic order in $\ta$, one finds
$$\tp = c +
\intinf\taumes \Bigl(\LpM \Bigr)^{{ik\over 2}}\,\tap +
\intinf\taumes (ik)\,\Bigl(\Lppp \Bigr) \,
\Bigl(\LpM\Bigr)^{{ik\over 2}-1}\, {\tap^2\over 2} + \CO(\tap^3)$$
where we have introduced
\eqn\ELpmdef
{\Lambda_{\pm} \equiv \pm \pp + x =
\pm M^{1/4}\,{\ee^{\pm 2\tau}\over \sqrt{\cosh 2\tau} }\,\,.}
Now using the fact that
$$\LpmM = 2\,M^{1/2}\,\ee^{\pm 2\tau}\,\,, $$
and
$$\Lpmpp = \pm {\ee^{\pm 2\tau} \over \sinh 2\tau}\,\,,$$
the above expansion becomes
\eqn\Etpexp
{\tp =(2\sqrt{M})^{ik/2}\,\Bigl[
c + \intinf\taumes \ektau \tap +
{ik\over 4\sqrt{M}}\,\intinf\taumes {\ektau\over \sinh 2\tau}\,
\tap^2 + \CO(\tap^3) \Bigr]\,\,.}
After an integration by parts, the above
expression in momentum space reads:
\eqn\Etpexpms
{\tp = (2\sqrt{M})^{ik/2}\,\Bigl[
c + \alpha_{-k} + k\int\d p_1 \int \d p_2 \,\,
{f(k+p_1+p_2)\over k+p_1+p_2-i\epsilon}\,\,
\alpha_{p_1} \alpha_{p_2} + \CO(\alpha^3) \Bigr]\,\,.}
When acting on the vacuum, this gives
precisely the result one finds for the connected contribution to an
out state in the lowest order of perturbation theory.
To obtain \Etpexpms\ use has been made of the fact that
$f(0)=0$. (Strictly speaking, this also implies
that the sign of the ${i\epsilon}$ prescription is not fixed to this
order. However, in the $c=1$ case, where $f(0)\ne 0$, a similar
analysis establishes the above operator as an out-state.)
A similar analysis can be carried out for $T^{(-)}_{-ik}$.

We will now provide further evidence that our interpretation of
the analytically continued operators $T^{(-)}_{-ik}$ ($\tp$) is correct
by deriving the tree-level scattering equation of ref.\refs{\rJeYo}
following the procedure developed in\refs{\rJRT}.
We first write $\tp$ as
$$\tp = \tpp - \tpm$$
where
\eqn\Etpp
{\tpp = \int \xmes \intap \da
\Bigl[ (\alpha +x)^2 + \mx \Bigr]^{ik/2}\,\,,}
and
\eqn\Etpm
{\tpm = \int \xmes \intam \da
\Bigl[ (\alpha +x)^2 + \mx \Bigr]^{ik/2}\,\,.}
After a change of variables to the ``time-of-flight'' variable
$\tau$, we will extend the limits of integration to the full line
in both the left and right sector. Dirichlet boundary conditions
then require
\eqn\Eexbc { \tpp = - \tpm \,\,. }

\noindent
Following Polchinski\refs{\rPo}, we consider the
asymptotic expansion
\eqn\Eashift
{\apm(x,t) = \pm x + {1\over x }\,\ttapm(\tau,t)+\CO({1 \over x^2})\,\,.}
For large $\tau$, we have
$$\pp(x)\to x\,, \qquad  x\to{M^{1/4}\over\sqrt{2}}\ee^{\tau}\,\,,$$
and therefore
\eqn\TEpm{\eqalign{
\tpm &= \intinf\taumes \int^{\ttam} \d\ta\,\,
\Bigl( {\ta^2 \over x^2} + {M \over x^2} \Bigr)^{{ik \over 2}}\cr
\noalign{\vskip 0.2truecm}
&= c + \intinf\taumes {M^{ik/2}\over x^{ik}} \sum_{p=0}^{\infty}
{1\over M^p}\,\,
{\Gamma (1+ ik/2) \over  p!\,(2p+1)\,\Gamma(1+ik/2-p)}
\,\,\ttam^{2p+1}\,\,.\cr }}
On the other hand
\eqn\TEpp{ \tpp = \intinf\taumes \int^{\ttap} \d\ta \,\,
\Bigl( 2M^{1/2}\,e^{2\tau} \Bigr)^{ik/2}
= c + \intinf\taumes (2M^{1/2})^{ik/2} \,
\ektau\, \ttap\,\,.}
Equation \Eexbc\ now implies
\eqn\EInOut{
- \ttap(\tau)= \sum_{p=0}^{\infty} {1\over M^p}\,\,
{ \Gamma (1 - \partial /2) \over
p!\,(2p+1)\,\Gamma(1-\partial /2-p)}\,\, \ttam^{2p+1}(-\tau)\,\,,}
in agreement with the solution of the scattering problem
obtained by Jevicki and Yoneya\refs{\rJeYo}. Expanding
\eqn\EExp{
\ttapm(\tau) = \intinf \d k\,\, \ttcpm_k\,\ee^{\pm ik\tau} }
the scattering equation \EInOut\ in momentum space becomes
\eqn\EInOutP{
- \ttcp_k = \sum_{p=0}^{\infty} {1 \over M^p}\,\,
{\Gamma (1 - ik/2) \over p!\,(2p+1)\,\Gamma (1-ik/2-p) }
\prod_{i=1}^{2p+1} \intinf
\d k_i \,\,\ttcm_{k_1}\ttcm_{k_2}\cdots\ttcm_{k_{2p+1}}
\delta(k-\sum k_i)\,\,.}
This equation contains the complete information about
tree-level scattering amplitudes and can be used to
obtain explicit expressions for arbitrary
$N\to M$ amplitudes. An analogous equation was derived for the
standard $c=1$ model in ref.\refs{\rMoPl}.
In the next section we shall describe an alternative
method to obtain scattering amplitudes using directly
the symmetry algebra. This method was developed for the
$c=1$ problem in \refs{\rJoao}.

\newsec{Scattering amplitudes}

We now proceed to obtain $1 \to N$ scattering amplitudes from the
analytically continued algebra \Ealg. The previous section
established, at least perturbatively,  the existence of a vacuum state
with respect to which
\eqn\Edefvacuum{
T^{(-)}_{-ik}\, |0\ket = |k;\,{\rm in}\,\ket\,, \qquad
\tp\, |0\ket = |k;\,{\rm out}\,\ket\,, \qquad
T^{(+)}_{-ik}\, |0\ket = 0\,\,. }
In order to compute a $1 \to N$ amplitude, it is  therefore
sufficient to obtain the ground state expectation value of the nested
commutator
\eqn\Ecomm{
\bra 0\vert\bigl[ T^{(+)}_{-ik_N} \ldots, \bigl[ T^{(+)}_{-ik_3}, \bigl[
\bigl[ T^{(+)}_{-ik_2}, \bigl[ T^{(-)}_{-ik_1} \bigr] \bigr] \ldots
\bigr]\vert 0\ket_c \,\,,}
where the subscript $c$ refers to connected diagrams.
Since the interaction part of the collective hamiltonian consists
of a cubic plus a tadpole term it is easy to see that any connected
diagram in a $1 \to N$ tree amplitude will involve exclusively
the cubic interaction (the tadpole term will contribute only to
disconnected diagrams).
Because the tadpole term is a quantum-mechanical effect resulting from the
normal ordering of the cubic interaction (see sect.~2),
for  tree-level amplitudes one can use the classical algebra \Ealg.
We will first evaluate
\eqn\BigComm{
\bigl[ T^{(+)}_{2\nn} \ldots, \bigl[ T^{(+)}_{2n_3}, \bigl[
\bigl[ T^{(+)}_{2n_2}, \bigl[ T^{(-)}_{2n_1} \bigr] \bigr] \ldots
\bigr]}
and then analytically continue the result and compute \Ecomm.

{}From the algebra \Ealg\ and the relationship \Ethreetwo\
one easily obtains the following useful results:
\eqn\Ecommonetwo{\eqalign{
\Bigl[ T^{(+)}_{2\nn}\,, O_{\snone - j,\snone - 2n_1}^{a=0} \Bigr]
&=  4\,i\,\nn\,(2\,n_1-j)\,\, O_{\sn -j-2,\sn -2n_1}^{a=1}\,\,,\cr
\noalign{\vskip 0.2truecm}
\Bigl[T^{(+)}_{2\nn}\,, O_{\snone - j,\snone - 2n_1}^{a=1} \Bigr]
&=4i\nn\,\,\Bigl( (2n_1+1-j)\,\, O_{\sn -j,\sn -2n_1}^{a=0} \cr
\noalign{\vskip 0.1truecm}
\qquad\qquad\qquad\qquad
&-4M(2n_1-j)\,\, O_{\sn -j-2,\sn -2n_1}^{a=0}\,\Bigr)\,\,,\cr}}
where $\Sigma_{{\rm L}} \equiv \sum_{i=1}^{L}\,n_i$.
It is then straightforward to evaluate the commutator \BigComm.
Denoting by $h_i \equiv (2\,n_1 - i)$ the result can be written as:
\eqn\Eexplicitcom{\eqalign{
N=2: \quad &2\,(4i)\,\, \bigl(\prod n_i\bigr)\,\,
O^{1}_{\sn-2,\sn-2n_1}\,\,,\cr
\noalign{\vskip 0.2truecm}
N=3: \quad &2\,(4i)^2 \bigl(\prod n_i\bigr)\,\Bigl[\,
h_1\,O^{0}_{\sn-2,\sn-2n_1}-4M h_2\, O^{0}_{\sn-4,\sn-2n_1}\,\Bigr]\,\,,\cr
\noalign{\vskip 0.2truecm}
N=4: \quad &2\,(4i)^3 \bigl(\prod n_i\bigr)\,\Bigl[\,
h_1 h_2\,O^{1}_{\sn-4,\sn-2n_1}
-4M h_2 h_4\, O^{1}_{\sn-6,\sn-2n_1}\,\Bigr]\,\,,\cr
\noalign{\vskip 0.2truecm}
N=5: \quad &2\,(4i)^4 \bigl(\prod n_i\bigr)\,\Bigl[\,
h_1 h_2 h_3\,O^{0}_{\sn-4,\sn-2n_1}
-4M h_2 h_4\,(h_1 + h_5)\, O^{0}_{\sn-6,\sn-2n_1} \cr
\noalign{\vskip 0.1truecm}
&\qquad\qquad\qquad\quad
+(4M)^2 h_2 h_4 h_6\, O^{0}_{\sn-8,\sn-2n_1}\,\Bigr]\,\,,\cr
\noalign{\vskip 0.2truecm}
N=6: \quad &2\,(4i)^5 \bigl(\prod n_i\bigr)\,\Bigl[\,
h_1 h_2 h_3 h_4\,O^{1}_{\sn-6,\sn-2n_1} \cr
\noalign{\vskip 0.1truecm}
&\qquad\qquad\qquad\quad
-4M h_2 h_4 h_6\,(h_1 + h_5 )\, O^{1}_{\sn-8,\sn-2n_1} \cr
\noalign{\vskip 0.1truecm}
&\qquad\qquad\qquad\quad
+(4M)^2 h_2 h_4 h_6 h_8\, O^{1}_{\sn-10,\sn-2n_1}\,\Bigr]\,\,,\cr
\noalign{\vskip 0.2truecm}
N=7: \quad &2\,(4i)^6 \bigl(\prod n_i\bigr)\,\Bigl[\,
h_1 h_2 h_3 h_4 h_5\,O^{0}_{\sn-6,\sn-2n_1}\cr
\noalign{\vskip 0.1truecm}
&\qquad\qquad\qquad\quad
-4M h_2 h_4 h_6\,(h_1 h_3+h_1 h_7+h_5 h_7)\,O^{0}_{\sn-8,\sn-2n_1}\cr
\noalign{\vskip 0.1truecm}
&\qquad\qquad\qquad\quad
+(4M)^2 h_2 h_4 h_6 h_8\,(h_1+h_5+h_9)\,O^{0}_{\sn-10,\sn-2n_1}\,\,,\cr
\noalign{\vskip 0.1truecm}
&\qquad\qquad\qquad\quad
-(4M)^3 h_2 h_4 h_6 h_8 h_{10}\, O^{0}_{\sn-12,\sn-2n_1}\,\Bigr]\,\,,\cr
\noalign{\vskip 0.2truecm}
N=8: \quad &2\,(4i)^7 \bigl(\prod n_i\bigr)\,\Bigl[\,
h_1 h_2 h_3 h_4 h_5 h_6\,O^{1}_{\sn-8,\sn-2n_1}\cr
\noalign{\vskip 0.1truecm}
&\qquad\qquad\qquad\quad
-4M h_2 h_4 h_6 h_8\,(h_1 h_3+h_1 h_7+h_5 h_7)\,O^{1}_{\sn-10,\sn-2n_1}\cr
\noalign{\vskip 0.1truecm}
&\qquad\qquad\qquad\quad
+(4M)^2 h_2 h_4 h_6 h_8 h_{10}\,(h_1+h_5+h_9)\,
O^{1}_{\sn-12,\sn-2n_1}\,\,,\cr
\noalign{\vskip 0.1truecm}
&\qquad\qquad\qquad\quad
-(4M)^3 h_2 h_4 h_6 h_8 h_{10} h_{12}\,
O^{1}_{\sn-14,\sn-2n_1}\,\Bigr]\,\,.\cr }}

We have also calculated nine- and  ten-point commutators.
The structure is now clear: for even (odd) $N$, we need to calculate
the $c$-number contribution to the ground state expectation value of
(analytically continued) operators $O^{a=0}\,\, (O^{a=1})$. Due
to energy conservation the commutators depend only on
the momentum of the single incoming particle, as is the case in the
standard $c=1$ model.

In order to calculate the $c$-number contribution to $O^{a=1}$
it is sufficient to obtain, say, $T^{\pm}_{2n}$ to linear
order in $\tap$. We find
$$ \eqalign {
T^{(+)}_{2n} &= c + 2^n M^{n/2} \intinf\taumes\,\, \ee^{2n\tau}\,\,
   \tap(\tau) + \CO(\tap^2)\,\,,\cr
\noalign{\vskip 0.2truecm}
 T^{(-)}_{2n} &= c + 2^n M^{n/2} \intinf\taumes\,\, \ee^{-2n\tau}\,\,
   \tap(\tau) + \CO(\tap^2)\,\,,\cr }
$$
and therefore
$$    [\,T^{(+)}_{2n_2},\,T^{(-)}_{2n_1}\,] =
     2in_1\,2^{n_1+n_2} M^{(n_1+n_2)/2} \intinf\taumes\,\,
     \ee^{2(n_2-n_1)\tau} + \CO(\tap)\,\,.
$$
Comparing this expression with what we expect from the algebra
\Ealg, we obtain
\eqn\Ecnum{
O_{2j,m}^{a=1}\Big\vert_{c-{\rm number}}
= { (4M)^{j+1} \over 4(j+1)} \delta_{m,0}\, , }
written in a form which anticipates the energy
conservation delta-function which is obtained once the analytical
continuation $n \to -ik/2$ is performed.

Tree-level amplitudes with an odd number of particles always
correspond to the $c$-number contribution to a commutator of the type
\eqn\Escom {[\, T^{(+)}_{2n} ,\, O_{j,m}^{a=1}\,]\,\,.}
We will now show that the operator $O_{j,m}^{a=1}$ has {\it no}
term linear in $\tap$.  We recall that
$$
O_{j,m}^{a=1} \equiv \int \xmes \intamp \da
\Bigl( \alpha^2 - x^2 +\mx \Bigr)
\Bigl( (\alpha + x)^2 +\mx \Bigr)^{{j+m\over 2}}
\Bigl( (\alpha - x)^2 +\mx \Bigr)^{{j-m\over 2}}\,\,.
$$
After shifting by $\pp$ the first term in brackets becomes
$${\ta^2 \over (\pi\phi_0)^2} + 2\ta +
  (\pi\phi_0)^2  - x^2 + {M \over x^2}\,\,.$$
The $\ta$-independent term vanishes because of \Ecsx\
and, apart from a $c$-number, $O_{j,m}^{a=1}$ starts
quadratically in $\tap$. Therefore, the commutator \Escom\
has no $c$-number, and all tree-level amplitudes for an odd number
of particles vanish. This is clearly related to
the fact that the Fermi energy of the problem is zero,
since otherwise the $\ta$-independent term above would be
$2\mf$, giving a nonvanishing $c$-number
contribution to \Escom.

Substituting eq. \Ecnum\ into eqs. \Eexplicitcom\
we obtain:
\eqn\Eamp{\eqalign{
N=2: \quad &(4i)\,\, \bigl(\prod n_i\bigr)\,(4M)^{n_1}
\Bigl({1\over 2n_1}\Bigr)\,\,
\delta_{\sn-2n_1,0}\,\,,\cr
\noalign{\vskip 0.2truecm}
N=4: \quad &(4i)^3 \bigl(\prod n_i\bigr)\,(4M)^{n_1}
\Bigl({1\over 4M}\Bigr)\,\,
\delta_{\sn-2n_1,0}\,\,,\cr
\noalign{\vskip 0.2truecm}
N=6: \quad &(4i)^5 \bigl(\prod n_i\bigr)\,(4M)^{n_1}
\Bigl[\, {3\,(2n_1-2) \over (4M)^2} \,\Bigr]\,\,
\delta_{\sn-2n_1,0}\,\,,\cr
\noalign{\vskip 0.2truecm}
N=8: \quad &(4i)^7 \bigl(\prod n_i\bigr)\,(4M)^{n_1}
\Bigl[\, {3\cdot5\,(2n_1-2)\,(2n_1-4) \over (4M)^3}\,\Bigr]\,\,
\delta_{\sn-2n_1,0}\,\,,\cr
\noalign{\vskip 0.2truecm}
N=10: \quad &(4i)^9 \bigl(\prod n_i\bigr)\,(4M)^{n_1}
\Bigl[\,{3\cdot5\cdot7\,(2n_1-2)(2n_1-4)\,(2n_1-6) \over (4M)^4}\,
\Bigr] \delta_{\sn-2n_1,0}\,\,.\cr}}
The structure of the amplitudes is now clear. Rescaling the exact
operators so that the coefficient of the linear term $\alpha_k$
equals one (see e.g. eq. \Etpexp), and analytically
continuing $n\to -ik/2$, we
finally obtain the general expression for $1 \to 2N-1$ amplitudes
($N \ge 1$):
\eqn\Eampl{
A(1\to 2N-1) = \Bigl( \prod_{i=1}^{2N} k_i\Bigr)\,\,
{ i\,(-)^{N-1} \over M^{N-1}} {1\over i\kn}
\Bigl[ (2N-3)!! \Bigl( \prod_{p=0}^{N-2} (i\kn +2p)\Bigr)\Bigr]
\delta (\sum_{i=1}^{2N-1} k_i - \kn)}
with all $1\to 2N$ amplitudes vanishing.
To conform with usual practice we have denoted by $\kn$ the momentum
of the single incoming particle.
We have confirmed this result by computing $2N-1 \to 1$ amplitudes
using the asymptotic in-out relationhsip
\EInOutP\ derived in sect.~3.

We close this section with a remark on scattering of special states.
If one assumes the existence of a vacuum state with respect to which
$T^{(-)}_{2n} |0\ket = \bra0|T^{(+)}_{2n} = 0$ (see, for instance,
\refs{\rGrDa}), it then follows  from our analysis
that the structure of $2N-1 \to 1$
amplitudes is identical to the one derived above.

\newsec{Quantum effects}

In this section we present some examples of loop
computations in quantum collective field theory. The
motivation for this is twofold: First, we want to
demonstrate the completeness at the quantum level of
the normal ordered  collective hamiltonian by
comparing our perturbative results with the corresponding
results obtained in the fermionic picture (and summarized
in appendix A). We believe that the
same procedure of normal ordering can then be
applied to the classical exact states discussed in
sect.~3 and will give exact quantum mechanical
scattering amplitudes.
Second,  there is certainly great interest in obtaining
quantum corrections to some of the results established
in\refs{\rJeYo} and in the previous sections.
These, in any case, still remain to be derived in the
conformal field theory approach.

We begin with the computation of the ground state energy.
The classical energy of the background field is
obtained from \Eezerodef\ and reads
\eqn\Eezero
{E^{(0)}=-{1\over 3\pi}\,\int \d x\,\,[-2v(x)]^{3/2}
=-{1\over 3\pi}\,\int_{M^{1/4}}^{\infty}\,\d x\,\,
\Bigl(x^2-\mx \Bigr)^{3/2}
= -{1\over 8\pi}\,M\ln M\,\,. }

The one-loop contribution is given by \Eeonedef
\eqn\Eeoneint
{E^{(1)}=\half\int\taumes \,\,\Delta(\tau)=
-{L\over 24\pi} +\,\,{\rm regular \,\, terms}\,\,,}
where $L$ is the extent of the $\tau$ coordinate
$$L=\int \d\tau = \int \,{\d x\over \pp} = -{1\over 4}\,\ln M\,\,,$$
giving
\eqn\Eeone
{E^{(1)} = {1\over 96\pi}\,\ln M\,\,.}

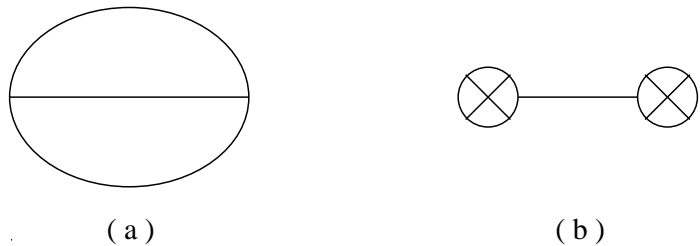
\ifig\fenergy{Ground state energy -- two-loop diagrams.}
{\epsfxsize3.0in\epsfbox{fig1.eps}}

The two-loop contribution is obtained from the second
order perturbation theory. There are two diagrams
shown in fig.~1 contributing:
\eqn\Eetwoint{\eqalign{
&E^{(2)}_a = -{1\over 6}\,\inthinf \d^3 k\,\,k_1 k_2 k_3\,\,
{f^2(k_1+k_2+k_3)\over k_1+k_2+k_3} =
{31\over 3780\pi}\, {1\over M}\,\,, \cr
\noalign{\vskip 0.3truecm}
&E^{(2)}_b = -{1\over 4}\,\inthinf \d k\,\, g^2(k) =
{1943\over 241920\pi}\, {1\over M}\,\,, \cr} }
which alltogether gives
\eqn\Eetwo
{E^{(2)}= {187\over 11520\pi}\,{1\over M}\,\,.}
In evaluating integrals we have used the $\zeta$-function
regularization as discussed in ref.\refs{\rDJR} and the
following results\refs{\rGR}:
\eqn\Eint{\eqalign{
&\inthinf \d x\,\, {x^{2m}\over \cosh^2 ax} =
{2^{2m}-2\over a}\,\Bigl({\pi\over 2a}\Bigr)^{2m}\,|B_{2m}|\,\,,
\quad m\ge 1\,\,, \cr
\noalign{\vskip 0.3truecm}
&\inthinf \d x\,x^{2m+1}\,{\sinh ax\over \cosh^2 ax} =
{2m+1\over a}\, \Bigl({\pi\over 2a}\Bigr)^{2m}\,|E_{2m}|\,\,,
\quad m\ge 0\,\,,\cr
\noalign{\vskip 0.3truecm}
&\inthinf {\d x\over \cosh^2 ax} = {1\over a}\,\,.\cr} }
$B_{2m}$ and $E_{2m}$ are Bernoulli and Euler numbers,
respectively.
Results \Eezero, \Eeone\ and \Eetwo\
are in precise agreement with the first three terms in the weak
coupling expansion of the exact result
given in appendix A.

\ifig\ftwopoint{First quantum correction to the two-point function.}
{\epsfxsize3.5in\epsfbox{fig2.eps}}

Next we evaluate the first quantum correction to the
two-point function using
\eqn\ESTdef
{S(k;k')=1-2\pi i \delta(k-k') T(k;k')\,\,.}
Again, there are two diagrams shown in fig.~2.
Their contributions are:
\eqn\Etpint{\eqalign{
&T_{2,a}^{(2)}(k)=-{k\over 12}\, \intinf \d p\,\,
p^3\,{f^2(p+k)\over p+k-i\epsilon \sign\,p} =
{1\over M}\,{1\over 360\pi}\,(28k + 15 k^3)\,\,,\cr
\noalign{\vskip 0.2cm}
&T_{2,b}^{(2)}(k)=-k \inthinf \d p\,\,f(p)\, g(p) =
{1\over M}\,{49k\over 720\pi} \,\,,\cr } }
which together give
\eqn\Etp
{T_2^{(2)}(k) = {1\over M}\,{1\over 48\pi}\,(7k+2k^3)\,\,.}

Finally, we discuss the finite temperature free energy.
The one-loop contribution
to the free energy is simply
\eqn\Efeone
{F^{(1)} =E^{(1)} + T\,{L\over \pi}\,\inthinf \d k\,\,
\ln(1-\ee^{-k/T} ) = {1\over 96\pi}\,(2\pi T)\,\,
\bigl[ {1\over 2\pi T} + 2\pi T \bigr]\,\ln M \,\,.}
The temperature dependent part above is just the free
energy of a massles boson on a half-line.
This is precisely the same answer (when expressed in
terms of the length $L$ of the spatial coordinate)
as in the standard $c=1$ case and simply means that
the two models cannot be distinguished at the
quadratic level. However, the order $\gst^2$ result
for the ground state energy is different for the
two models. With this in mind we next compute the
two-loop result for the free energy which contains
physical information about degrees of freedom.

The two-loop contribution to the free energy
is easily computed once we know the second order
two- and four-point functions, as was done in\refs{\rDJR}
using the result of ref.\refs{\rDaMa}:
\eqn\Efetwoint{
F^{(2)}=E^{(2)}+\inthinf \d k\,n(k)\,\Re T_2^{(2)}(k;k)+
\inthinf \d k_1 \inthinf \d k_2\, n(k_1) n(k_2)\,
\Re T_4^{(2)}(k_1,k_2;k_1,k_2) }
where $n(k)=1/(\ee^{k/T}-1)$. Using the result \Etp\
for $T_2^{(2)}(k)$ and the expression for the four-point
amplitude derived in\refs{\rJeYo}:
\eqn\Efp
{T_4^{(2)}(k_1,k_2;k_1,k_2)={1\over M}\,{1\over 4\pi}\,k_1 k_2\,\,,}
we get
\eqn\Efetwo
{F^{(2)}={1\over M}\,\,{1\over 11520\pi}\,\,{(2\pi T)^2}\,\,
\Bigl[\,{187\over (2\pi T)^2}+70 + 7\,(2 \pi T)^2 \Bigr] \,\,.}
As a check, equation \Efetwo\ is derived in the fermionic
description of the model at the end of appendix A.

The beauty of the formula \Efetwoint\ is that it allows one
to easily see which states in the spectrum contribute to
the free energy at this order. Namely, one
simply finds from the integral representation of scattering
amplitudes which states in the intermediate channel
give rise to the real part of the amplitude.
In our case the contribution from the exchange of
continuous mode (which would be imaginary) vanishes
since it is proportional to $f(0)$, and the total answers
for the second order two- and four-point amplitudes
come entirely from an exchange of states at discrete
values of momenta discussed at the end of sect.~2.
There is, however, a puzzle concerning result \Efetwo\
in that it is not symmetric under the duality transformation
\eqn\Eduality{
2\pi T \to {1 \over 2\pi T}\,\,.}
If it were,
the expression in the brackets in \Efetwo\ would be symmetric under
\Eduality.  We do not yet have a complete understanding of this.
However, we believe this to be related to the fact that the
collective field amplitudes used in \Efetwoint\ are
only a part of the total tachyon amplitudes. If the
factorization assumption is correct, these amplitudes
have to be multiplied by external leg factors as discussed
in the introduction. Since one expects the string theory
free energy to be symmetric under the duality transformation,
our result would imply that the external
leg factors (in Minkowski space) must have such a specific momentum
dependence that when the full amplitudes are integrated
according to formula \Efetwoint\ the obtained answer is symmetric
under \Eduality. This observation may be useful in finding
the correct completion of the collective field amplitudes
to the full answer. We note that in the standard
$c=1$ case  collective field theory gave the
full (dual) answer for the free energy up to second
order\refs{\rDJR}, which could be related to the fact that external
leg factors in this case are pure phases.

\newsec{Discussion and conclusions}

We have investigated further the collective
field theory description of a deformed matrix model with
emphasis on states, symmetries and higher-loop corrections.
We have identified the exact operators which create in
and out tachyon states. By imposing an appropriate boundary
condition we showed that these states encode the complete
information on the tree-level scattering amplitudes which
can be written in a form of a scattering equation\refs{\rJeYo}.
We then proceeded to derive amplitudes using directly
(analytic continuation of) the symmetry algebra.
In both computations it was clear that vanishing of the
``odd-point'' on-shell amplitudes is closely related to the
choice of zero Fermi energy. In the collective hamiltonian
this choice is seen as the vanishing
of the three-string vertex at zero momentum.

We computed loop corrections to the ground state
energy, finite temperature free energy and the
two-point function, and found the interesting result that the
free energy is not invariant under the duality transformation
typical of compactified string theory.
This result may be helpful for a better understanding
of the full tachyon amplitudes in the black hole background.

We hope we have further clarified some issues and pointed out
to some problems in the matrix model approach to the black
hole solution of two-dimensional string theory.
In particular, we believe that our discussion of
classical exact states can be extended to the quantum
level by normal ordering as explained in sect.~2. This would
then give the full quantum collective field amplitudes
and allow one to check the consistency of the factorization
assumption at the loop-level. The exact collective field
amplitudes can probably be obtained in the fermionic
picture of matrix model following\refs{\rMoore},
and work in this direction is in progress.
Finally, any explicit result, either for amplitudes or
for the partition function\refs{\rBeKl} obtained in
the continuum approach would greatly help to improve
our understanding of a possible matrix model formulation
of the black hole problem.
It is precisely this interplay between the two
approaches which has lead to remarkable progress in our
understanding of the standard $c=1$ problem, and we
certainly hope this to continue.

\bigskip
\noindent
{\bf Note added}

During the completion of this manuscript we have
learned of the work of Danielsson (ref.\refs{\rDa})
who derived some of the results presented in sect.~5
using different methods from ours.

\bigskip
\noindent
{\bf Acknowledgments}

We wish to thank Antal Jevicki and Igor Klebanov for
discussions. Part of this work was done while one of us
(K.D.) was visiting the Centre for Nonlinear Studies
and Physics Department
of the University of Witwatersrand in Johannesburg.
The warm hospitality extended to him during
the visit is greatfully acknowledeged.
The work of K.D. is supported in part by NSF Presidential
Young Investigator Award PHY-9157482 and James S.
McDonnell Foundation grant No. 91-48.

\bigskip
\appendix{A}{}

In this appendix we derive an expansion for the
ground state energy directly in the fermionic picture of
the matrix model. It is given by
\eqn\Eedef
{E=\int^{\mf} \d e\,\, e\rho(e)}
where $\rho(e)$ is the density of states, and in our
case $\mf=0$. In the standard $c=1$ model one
had $\mf = -\mu \not= 0$ and the problem could
be solved by knowing only the density of states at the
Fermi level\refs{\rGrMi}. A similar trick can be used here, 
but now one needs
to know the density of states as a function of energy, and
only at the end can one put $e=0$.

The density of states is given by
\eqn\Erhoedef
{\rho(e) ={1\over \pi}\,\,\Im\,\sum_{n=0}^{\infty}\,\,
{1\over \epsilon_n - e -i\epsilon} }
where $\epsilon_n$ are the single particle energy levels
which can be obtained by solving  the Schr\"odinger
equation with potential \Edmm.
In ref.\refs{\rLa} the eigenvalue problem for
a ``deformed harmonic potential''
\eqn\Epotential{
v(x) = \half \omega^2 x^2 +{\lambda\over x^2}}
was solved with the result
\eqn\Eenergies{
\epsilon_n =(2n+1+a)\,\omega\,,\qquad\quad
a\equiv \half\,\sqrt{1+8\lambda}\,\,.}
Strictly speaking, there is a second set of wave functions with
energies $\epsilon_n =(2n+1-a)\,\omega$, which although singular at the
origin, are also normalizable on the half-line. Both branches can be
extended to the full line, the first being chosen to be odd and
the second even under parity\refs{\rLa}. As $\lambda \to 0$, these
wave functions and corresponding eigenvalues map smoothly into the
odd (respectively even) solutions of the harmonic oscillator problem.
However, in the first two references of\refs{\rLa}
it is argued that the second
set of wave functions is not physically acceptable because, for instance,
the laplacian operator would not be hermitian. For us, not to consider this
set is a reasonable choice for at least two reasons: first, we
only consider the right branch of the potential and the first
set of wave functions indeed vanishes at the origin; second,
as $\lambda \to 0$, we only recover half the energies of the
harmonic oscillator. This is a desirable feature following
our previous observation that for the inverted case, the full algebra
of the deformed matrix model reduces to ``half'' the
discrete state $W_{\infty}$ algebra of the $M\to 0$ case.

Having justified our choice of energies, we proceed with
evaluation of the density of states \Erhoedef.
For our problem with an inverted harmonic
potential we have to analytically continue $\omega\to -i\omega$
(and set $\omega = 1$)  which gives
\eqn\Erhoestart
{\rho(e) =-{1\over 2\pi}\, \Re\,\sum_{n=0}^{\infty}\,\,
{1\over n+\half+{a-ie\over 2} } \,,\qquad\quad
a=\sqrt{M+1/4}\,\,.}
Expanding this as in ref.\refs{\rGrMi} one finds
\eqn\Erhoe
{\rho(e)=-{1\over 2\pi}\,\Re\Bigl[ \ln (a-ie) +
\sum_{m=1}^{\infty}\,\,
{(2^{2m-1} - 1)\,B_{2m}\over m}\,\,
{1\over (a-ie)^{2m}}\,\Bigr]\,\,.}
The first few terms in this expansion read
\eqn\Erhoeexp
{\rho(e)=-{1\over 4\pi}\,\ln(a^2+e^2) -
{1\over 2\pi}\,\Bigl[ {1\over 6}\,{a^2-e^2\over(a^2+e^2)^2} -
{7\over 60}\,{a^4-6a^2 e^2 +e^4\over(a^2+e^2)^4} + \cdots \Bigr]\,\,.}
One could now simply evaluate integral in \Eedef\ using
\Erhoeexp\ term by term. However, one can avoid this
procedure by noting that the energy given by \Eedef\
corresponds to the solution of the equation
\eqn\Eediffeq{
\rho(e)=-\,{\partial^2 E(e)\over \partial e^2} }
for $e=0$. One easily finds that \Eediffeq\ is solved by
\eqn\Esol
{E(e)={1\over 2\pi}\Re\Bigl[-\half(a-ie)^2\ln (a-ie) +
{1\over 6}\ln (a-ie) - \sum_{m=2}^{\infty}
{(2^{2m-1} - 1)\,B_{2m} \over m(2m-2)(2m-1)}
{1\over (a-ie)^{2m-2}}\Bigr] }
from which it follows that
\eqn\Eexact
{E= -{1\over 2\pi}\,\Bigl[\half a^2\ln a - {1\over 6}\,\ln a
+\sum_{m=1}^{\infty}\,
{(2^{2m+1} - 1)\,B_{2m+2} \over 2m(m+1)(2m+1)}\,\,
{1\over a^{2m}}\,\Bigr]\,\,. }
However, we are interested in an expansion in
$\gst^2 = 1/M$, and therefore one still needs to reexpand \Eexact\
in powers of $1/M$ using that $a=\sqrt{M+ 1/4}$.
This gives
\eqn\Eexactexp
{E=-{1\over 8\pi}\,M \ln M + {1\over 96\pi}\,\ln M
+\sum_{n=1}^{\infty}\, {a_{n}\over M^n}\,\,,}
with the coefficients $a_n$ given by
\eqn\Eecoeff
{a_n = {1\over 2\pi}{(-1)^{n+1}\over 4^n}\Bigl[
{4n+1\over 48n(n+1)} +
\sum_{m=1}^{n}\, {(-1)^m \,4^m\,(2^{2m+1}-1)\,B_{2m+2} \over
2m(m+1)(2m+1)}\,
{(n-1)!\over (m-1)! (n-m)!}\Bigr]\,\,.}
The first few terms in this expansion are:
\eqn\Eeexp{
E=-{1\over 8\pi}\,M \ln M + {1\over 96\pi}\,\ln M
+{187\over 11520\pi}\, {1\over M}
-{3083\over 322560\pi}\,{1\over M^2}
+{4719\over 286720\pi}\,{1\over M^3} - \cdots\,\,.}

We have also checked that the analytic continuation we have
done is justified by computing the first few terms in the expansion
of $\rho(e)$ using the Gelfand-Dikii formula\refs{\rGeDi,\rGrMi}
$$\rho(e)= {2\over\pi}\,\int \d y\,\,\sum_{\ell=0}^{\infty}\,
(-1)^{\ell}\,{R_{\ell}\,[2(v(x) - v(y))] \over
[2(e-v(y))]^{\ell+\half} } \Bigg\vert_{x=y} $$
where $R_{\ell}[u]$ are Gelfand-Dikii polynomials and
the integral goes over the interval where $e-v(y)\ge 0$.
We have found precise agreement with \Erhoeexp.
This is to be expected since
the expressions for the energy \Eetwoint, when
expressed as $x$-space integrals, are in agreement with the Gelfand-Dikii
expansion for the density of states for an {\it arbitrary} potential,
once use has been made of eq. \Eediffeq,
as it was pointed out in the first of refs.\refs{\rDJR}.

Finally, we summarize the computation of the finite
temperature free energy in the fermionic picture.
It can be written as\refs{\rGrKlcircle}:
\eqn\Efedef{
F=\int \d e\,E(e)\,{\partial\over\partial\mf}\,
\left( {1\over 1+\ee^{(e-\mf)/T}}\,\right) =
\int \d e\,E(e)\,\,{1\over 4T\,\cosh^2 
\bigl({e-\mf\over 2T}\bigr)}\,\,,}
where $E(e)$ is defined in \Eedef.
Integrating by parts we obtain
$$F= -\int \d e\,\rho(e)\,\,{\partial^2\over\partial e^2}\,\,
{1\over 4T\,\cosh^2 \bigl({e-\mf\over 2T}\bigr)}=
-{\partial^2\over \partial\mf^2}\,\int\d e\,
{\rho(e)\over 4T\,\cosh^2 \bigl({e-\mf\over 2T}\bigr)}\equiv
-{\partial^2\over \partial\mf^2}\,\tilde\rho(\mf)\,\,.$$
It is straightforward to check that for the
standard $c=1$ model the above equation is in agreement with
the relationship between $F$ and $\tilde\rho$ derived
in\refs{\rGrKlcircle}.
{}From eqs. \Efedef\ and \Esol, we obtain (after setting
$\mf=0$):
\eqn\Efeintrep{\eqalign{
F=&-{1\over 2\pi}\,\int\d e\,\,
\Re\Bigl[\,\half(a-ie)^2\,\ln (a-ie) -B_2\,\ln (a-ie) \cr
\noalign{\vskip 0.2truecm}
&+ \sum_{m=2}^{\infty} {(2^{2m-1} - 1)\,B_{2m} \over m(2m-2)(2m-1)}
\,\,{1\over (a-ie)^{2m-2}}\, \Bigr]\,\,
{1\over 4T\,\cosh^2 \bigl({e\over 2T}\bigr)} \,\,.\cr }}
Expanding, the first few terms in powers of $1/M$ are:
\eqn\Efeexpansion{\eqalign{
F=&-{1\over 2\pi}\,\biggl\{\,
{1\over 4}\,I_0 M\ln M + \Bigl[\bigl({1\over 16}-{B_2\over 2}\bigr)I_0
-{1\over 4}\,I_2\Bigr]\,\ln M \cr
\noalign{\vskip 0.2truecm}
&+\Bigl[\bigl({1\over 128}-{B_2\over 8} +{7B_4\over 12}\bigr)\,I_0
-\bigl({1\over 16} +{B_2\over 2}\bigr)\,I_2 -{1\over 24}\,I_4
\Bigr]\,{1\over M}+ \dots \biggr\}\,\,,\cr}}
where
$$I_{2n} = \intinf \d x\,\,
{x^{2n}\over 4T\,\cosh^2 \bigl({e\over 2T}\bigr)}
=\bigl(1-2^{1-2n}\,\bigr)\,\vert B_{2n}\vert\,(2\pi T)^{2n}\,,
\quad n=1,2,3,\dots$$
and $I_0=1$. This finally gives
\eqn\Efefinal{
F= -{1\over 8\pi}\,M \ln M + {1\over 96\pi}\,
\bigl[(1+(2\pi T)^2\bigr]\,\ln M
+{1\over M}\,\,
{1\over 11520\pi}\,\bigl[ 187 + 70(2\pi T)^2 + 7(2\pi T)^4\bigr]\,\,,}
in precise agreement with our results of sect.~5.

\listrefs
\vfill\eject
\bye